\documentclass[preprint,showpacs,preprintnumbers,amsmath,amssymb]{revtex4}

\usepackage{graphicx}
\usepackage{dcolumn}
\usepackage{bm}

\begin{document}

\preprint{APS/123-QED}

\title{ Effect of isospin dependent cross-section on fragment production in the collision of charge asymmetric nuclei\\}
\author{Anupriya Jain}
\author{Suneel Kumar}%
 \email{suneel.kumar@thapar.edu}
\affiliation{%
School of Physics and Material Science, Thapar University, Patiala-147004, Punjab (India)\\
}%
\date{\today}
\begin{abstract}
To understand the role of isospin effects on fragmentation due to the collisions of charge asymmetric nuclei, we have performed a complete systematical study using isospin dependent quantum molecular dynamics model. Here simulations have been carried out for $^{124}X_{n}+ ^{124}X_{n}$, where n varies from 47 to 59 and for $^{40}Y_{m}+ ^{40}Y_{m}$, where m varies from 14 to 23. Our study shows that isospin dependent cross-section shows its influence on fragmentation in the collision of neutron rich nuclei.\\
\end{abstract}
\maketitle
\baselineskip=1.5\baselineskip\
\section{Introduction}
Heavy-ion collisions have been extensively studied over the last decades.
The behavior of nuclear matter under the extreme conditions of temperature, density, angular momentum etc., is a very important aspect of heavy-ion physics. Multifragmentation is one of the extensively studied field at intermediate energies. One of the major ingredient in heavy ion collisions is the symmetry energy, whose form and strength is one of the hot topic these days \cite{1}. This quantity vanishes at a certain incident energy. Finite nuclei studies predict values for the symmetry energy at saturation of the order of 30-35 MeV. In heavy ion collisions highly compressed matter can be formed for short time scales, thus the study of such a dynamical process can provide useful information on the high density dependence of symmetry energy. Even at low incident energies which belong to even smaller baryonic densities, the isospin dependence of the mean field potential was shown to yield same result obtained with potentials that has no isospin dependences. These results are in similar lines and it also indicates that even binary phenomena like fission will also be insensitive towards isospin dependence of the dynamics \cite{2}. Recently theoretical studies on the high density symmetry energy have been started by investigating heavy ion collisions of asymmetric systems \cite{3,4}. Comparisons of collisions of neutron-rich to that of neutron-deficient systems provide a means of probing the asymmetry term experimentally \cite{5,6,7}. The experimental analysis of the isospin effects on fragment production has yielded several interesting observations: Dempsey et al. \cite{8} in their investigation of $^{124,136}Xe+^{112,124}Sn $ at 55 MeV/nucleon found that multiplicity of IMF's increases with the neutron excess of the system. A more comprehensive study was carried out by Buyukcizmeci et al. \cite{9} showed that symmetry energy  of the hot fragments produced in the statistical freeze-out is very important for isotope distributions, but its influence is not very large on the mean fragment mass distributions.  Symmetry energy effect on isotope distributions can survive after secondary de-excitation. Moreover Schmidt et al. \cite{10} in their investigation on the analysis of LCP's production and isospin dependence of $^{124}Sn+^{64}Ni$, $^{124}Sn+^{58}Ni$, $^{124}Sn+^{27}Al$ at 35 MeV/nucleon and 25 MeV/nucleon collisions found that isospin effects were demonstrated in the observables, such as the angular distribution of light particles emitted in central collisions at 35 MeV/nucleon and LCP's emission. On the other hand Tsang et al. \cite{11} in their investigation of $^{112}Sn + ^{124}Sn$, $^{124}Sn + ^{112}$Sn systems at an incident energy of E=50 MeV/nucleon showed the effects of isospin diffusion by investigating heavy-ion collisions with comparable diffusion and collision time scales. They showed that the isospin diffusion reflects driving forces arising from the asymmetry term of the EOS. With the passage of time, isospin degree of freedom in terms of symmetry energy and nucleon-nucleon cross section is found to affect the balance energy or energy of vanishing flow and related phenomenon in heavy-ion collisions \cite{12}. J. liu et al.,\cite{13} studied the effect of Coulomb interaction and symmetry potential on the isospin fragmentation ratio $(N/Z)_{gas}/(N/Z)_{liq}$ and nuclear stopping R. They showed that Coulomb interaction induces important isospin effects on both $(N/Z)_{gas}/(N/Z)_{liq}$ and R.  However, the isospin effects of symmetry potential and Coulomb interaction on $(N/Z)_{gas}/(N/Z)_{liq}$ and R are different.\\
Our present study will shed light on isospin effects on multiplicity of fragments produced in the collision of charge asymmetric colliding nuclei. We present microscopic study of isospin dependent nucleon-nucleon cross section on charge asymmetric nuclear matter. In this paper our aim is two fold, one is to look for effect of density dependent symmetry energy on fragmentation and second is to look for influence of isospin dependent and isospin independent cross-section on fragmentation due to collision of charge asymmetric nuclei.\\
This study is carried out within the framework of isospin-dependent quantum molecular dynamics model
that is explained in section-II. The results are presented in section-III. We present summary in section-IV.\\

\section{ISOSPIN-DEPENDENT QUANTUM MOLECULAR DYNAMICS (IQMD) MODEL}

Theoretically many models have been developed to study the heavy ion collisions at intermediate energies. One of them is quantum molecular dynamical model (QMD) \cite{14,15}, which incorporates N-body correlations as well as nuclear EOS along with important quantum features like Pauli blocking and particle production.\\
In past decade, several refinements and improvements were made over the original QMD \cite{14,15}. The IQMD \cite{16} model overcomes the difficulty as it not only describe the ground state properties of individual nuclei at initial time but also their time evolution. In order to explain experimental results in much better way and to describe the isospin effect appropriately, the original version of QMD model was improved which is known as isospin-dependent quantum molecular dynamics (IQMD) model.\\
The isospin-dependent quantum molecular dynamics (IQMD)\cite{16} model treats different charge states of
nucleons, deltas and pions explicitly, as inherited from the VUU model. The IQMD model has been used successfully
for the analysis of large number of observables from low to relativistic energies. The isospin degree of
freedom enters into the calculations via both cross-sections and mean field.\\
In this model,baryons are represented by Gaussian-shaped density distributions
\begin{equation}
f_i(\vec{r},\vec{p},t) = \frac{1}{\pi^2\hbar^2}~e^{-(\vec{r}-\vec{r_i}(t))^{2}\frac{1}{2L}}~e^{-(\vec{p}-\vec{p_i}(t))^{2}\frac{2L}{\hbar^2}}.
\end{equation}
Nucleons are initialized in a sphere with radius $R= 1.12 A^{1/3}$ fm, in accordance with the liquid drop model. Each nucleon occupies a volume of $h^3$, so that phase space is uniformly filled. The initial momenta are randomly chosen between 0 and Fermi momentum($p_F$). The nucleons of target and projectile interact via two and three-body Skyrme forces and Yukawa potential. The isospin degree of freedom is treated explicitly by employing a symmetry potential and explicit Coulomb forces
between protons of colliding target and projectile. This helps in achieving correct distribution of protons and neutrons
within nucleus.\\
The hadrons propagate using Hamilton equations of motion:
\begin{equation}
\frac{d{r_i}}{dt}~=~\frac{d\it{\langle~H~\rangle}}{d{p_i}}~~;~~\frac{d{p_i}}{dt}~=~-\frac{d\it{\langle~H~\rangle}}{d{r_i}},
\end{equation}
with
\begin{eqnarray}
\langle~H~\rangle&=&\langle~T~\rangle+\langle~V~\rangle\nonumber\\
&=&\sum_{i}\frac{p_i^2}{2m_i}+
\sum_i \sum_{j > i}\int f_{i}(\vec{r},\vec{p},t)V^{\it ij}({\vec{r}^\prime,\vec{r}})\nonumber\\
& &\times f_j(\vec{r}^\prime,\vec{p}^\prime,t)d\vec{r}d\vec{r}^\prime d\vec{p}d\vec{p}^\prime .
\end{eqnarray}
 The baryon-baryon potential $V^{ij}$, in the above relation, reads as:
\begin{eqnarray}
V^{ij}(\vec{r}^\prime -\vec{r})&=&V^{ij}_{Skyrme}+V^{ij}_{Yukawa}+V^{ij}_{Coul}+V^{ij}_{sym}\nonumber\\
&=& \left [t_{1}\delta(\vec{r}^\prime -\vec{r})+t_{2}\delta(\vec{r}^\prime -\vec{r})\rho^{\gamma-1}
\left(\frac{\vec{r}^\prime +\vec{r}}{2}\right) \right]\nonumber\\
& & +~t_{3}\frac{exp(|\vec{r}^\prime-\vec{r}|/\mu)}{(|\vec{r}^\prime-\vec{r}|/\mu)}~+~\frac{Z_{i}Z_{j}e^{2}}{|\vec{r}^\prime -\vec{r}|}\nonumber\\
& & + t_{6}\frac{1}{\varrho_0}T_3^{i}T_3^{j}\delta(\vec{r_i}^\prime -\vec{r_j}).
\label{s1}
\end{eqnarray}
Here $Z_i$ and $Z_j$ denote the charges of $i^{th}$ and $j^{th}$ baryon, and $T_3^i$, $T_3^j$ are their respective $T_3$
components (i.e. 1/2 for protons and -1/2 for neutrons). Meson potential consists of Coulomb interaction only.
The parameters $\mu$ and $t_1,.....,t_6$ are adjusted to the real part of the nucleonic optical potential. For the density
dependence of nucleon optical potential, standard Skyrme-type parameterization is employed.
The choice of equation of state (or compressibility) is still controversial one. Many studies
advocate softer matter, whereas, much more believe the matter to be harder in nature. 
We shall use soft (S) equation of state that have compressibility of 200 MeV.\\

The binary nucleon-nucleon collisions are included by employing the collision 
term of well known VUU-BUU equation. The binary collisions
are done stochastically, in a similar way as are done in all transport models. During the propagation, two nucleons are
supposed to suffer a binary collision if the distance between their centroids
\begin{equation}
|r_i-r_j| \le \sqrt{\frac{\sigma_{tot}}{\pi}}, \sigma_{tot} = \sigma(\sqrt{s}, type),
\end{equation}
"type" denotes the ingoing collision partners (N-N, N-$\Delta$, N-$\pi$,..). In addition,
Pauli blocking (of the final
state) of baryons is taken into account by checking the phase space densities in the final states.
The final phase space fractions $P_1$ and $P_2$ which are already occupied by other nucleons are determined for each
of the scattering baryons. The collision is then blocked with probability
\begin{equation}
P_{block}~=~1-(1-P_1)(1-P_2).
\end{equation}
The delta decays are checked in an analogous fashion with respect to the phase space of the resulting nucleons.\\

\section{Results and Discussion}
To check the influence of density dependent symmetry energy on fragmentation we have simulated $_{50}Sn^{124}+_{50}Sn^{124}$  and $_{50}Sn^{107}+_{50}Sn^{124}$   reactions by using isospin dependent quantum molecular dynamics (IQMD) model  at incident energy 600 MeV/nucleon for complete colliding geometry. The phase space generated using IQMD model has been analyzed using minimum spanning tree [MST] algorithm \cite{16} and minimum spanning tree with momentum cut \cite{17} [MSTP]. The results obtained are discussed as follows: \\
\begin{figure}
\includegraphics[scale=0.38]{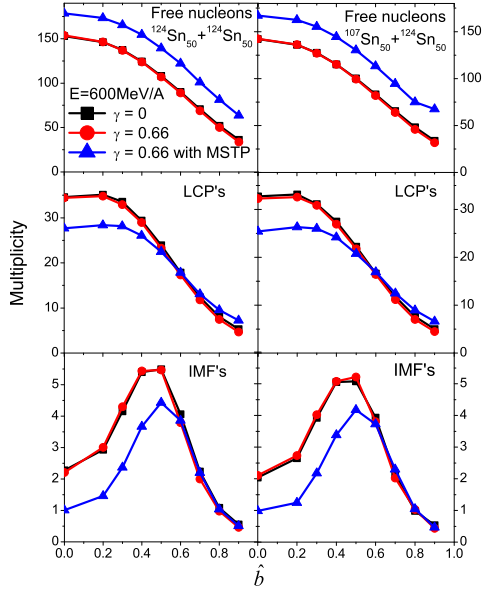}
\caption{\label{Fig:1} Multiplicity of free nucleons, LMF's and IMF's as a function of scaled impact parameter.}
\end{figure}
 In Fig.\ref{Fig:1}, shows multiplicity of free nucleons, LMF's and IMF's as a function of scaled impact parameters for 
 $_{50}Sn^{124}+_{50}Sn^{124}$  and $_{50}Sn^{107}+_{50}Sn^{124}$ . 
Our findings are as follows:\\ 
(1) As we move from central to peripheral collisions the number of free nucleons and LMF's decreases because the participation zone decreases which leads to the lower number of free nucleons and LMF's. But in case of IMF's, curve shows a "rise and fall" this is because for central collision the overlapping of participant and spectator zone is maximum so we get very small number 
of IMF's. For semi peripheral collisions the participant and spectator zone decreases so the production of IMF's increases and for peripheral collisions very small portion of target and projectile overlap so again few number of IMF's observed most of the fragments goes out as heavy mass fragments (HMF's).\\ 
(2) Number of free nucleons, LMF's and IMF's in $_{50}Sn^{107}+_{50}Sn^{124}$ produced are smaller as compare to $_{50}Sn^{124}+_{50}Sn^{124}$. This is because for the neutron rich system, heavy residue with low excitation energy will predominantly emit neutrons, a channel that is suppressed in case of neutron poor nuclei.\\ The equation\\
\begin{equation}
E(\rho)= E(\rho_{0})(\frac{\rho}{\rho_{0}})^\gamma
\end{equation}
 gives us the theoretical conjecture of how symmetry energy varies against density. $\gamma$, tells us the stiffness of the symmetry energy \cite{18}. In Fig.1, the free nucleons, LMF's and IMF's for $\gamma= 0$ and $\gamma= 0.66$ both the curves clearly indicates the density dependence of symmetry energy. From fig.1, one can see that:\\
(1) Small difference is observed in both curves in case of LMF's at a scaled impact parameter range from 0.0 to 0.4. This is because the density during the fragmentation is smaller than normal nuclear matter density. Hence the role of density dependent symmetry energy is negligible.\\ 
(2)  When we apply momentum cut in addition to space cut the number of free nucleons increases because at low impact parameter participant zone increases and large number of free nucleons produced on the other hand number of LMF's and IMF's decreases with MSTP cut.\\
 Although we have tried the simulation for two different parameterization of density dependent symmetry energy but influence of this is very small, because the density at which fragmentation take place is lower than normal nuclear matter density. Hence influence of symmetry energy on fragmentation is very small, which is in agreement with the observation of ref. \cite{19}.\\ 
\begin{figure}
\includegraphics[scale=0.38]{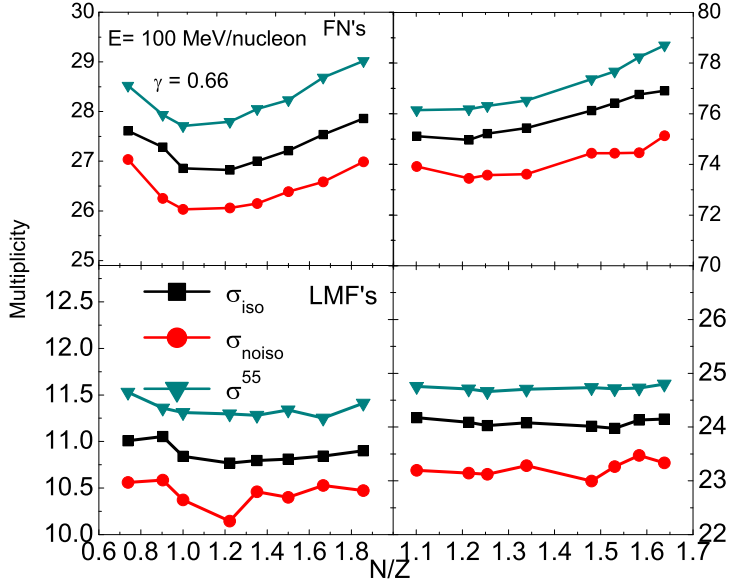}
\caption{\label{Fig:2} Multiplicity of free nucleons and LMF's with N/Z.}
\end{figure}
Now to check the role of different cross-sections on fragmentation for charge asymmetric colliding nuclei, we have chosen two set of reactions, one where mass of colliding nuclei is 40, but charge varies from 14 to 23. For first set the chosen reactions are $^{40}X_{m}+^{40}X_{m}$, where $^{40}X_{m}$ = ($^{40}V_{23}$, $^{40}Sc_{21}$, $^{40}Ca_{20}$, $^{40}Ar_{18}$, $^{40}Cl_{17}$, $^{40}S_{16}$, $^{40}P_{15}$ and $^{40}Si_{14}$ ) respectively. For second set we have chosen the reactions for which mass of colliding nuclei is 124, but charge varies from 47 to 59. Second set of reactions taken are $^{124}Y_{n}+^{124}Y_{n}$, where $^{124}Y_{n}$ = ($^{124}Ag_{47}$, $^{124}Cd_{48}$,  $^{124}In_{49}$, $^{124}Sn_{50}$, $^{124}I_{53}$, $^{124}Cs_{55}$, $^{124}Ba_{56}$ and $^{124}Pr_{59}$ ) respectively. All the simulations are carried out for $\hat {b}$ = 0.3 at 100 MeV/nucleon 
for symmetry energy corresponding to $\gamma= 0.66$. Here we take three different nucleon-nucleon cross-section because at low energy cross-section have very large influence on fragment production. Moreover, they have small effect on fragment production for central collision, whereas fragment production is strongly influenced at semicentral ($\hat {b}$=0.3 in this case) collisions \cite{20}.
 In Fig.\ref{Fig:2}, we have displayed the multiplicity of free nucleons (FN's) and light mass fragments (LMF's) as a function of charge asymmetry (N/Z). One can clearly see the effect of different cross-section on the production of FN's and LMF's. It has been observed that:\\ 
(1) If we fix $\sigma_{nn}$ = $\sigma_{pp}$ = $\sigma_{np}$ = 55mb, then maximum production of FN's and LMF's takes place. Isospin effect can be clearly seen when we use isospin dependent cross-section $\sigma_{iso}$ ($\sigma_{np}$ = 3$\sigma_{nn}$ = 3$\sigma_{pp}$) and isospin independent cross-section $\sigma_{noiso}$ ($\sigma_{nn}$ = $\sigma_{pp}$ = $\sigma_{np}$).\\
 $\sigma_{noiso}$ will reduce the cross-section and thus the number of collisions, hence lead to less production of FN's and LMF's but $\sigma_{iso}$ will enhance the number of collisions and hence the production of FN's and LMF's. Moreover, one can see that nearly constant difference in the production with $\sigma_{iso}$ and $\sigma_{noiso}$.\\ 
(2) Minimum production takes place for the case when N/Z = 1 i.e symmetric charge collisions. As we know that nuclei offer very interesting isospin situation where, the symmetry potential, Coulomb interaction and isospin dependent nucleon-nucleon collisions are simultaneously present. The Coulomb interaction is an important asymmetry term which can bring an important isospin effect into the observable quantities in the intermediate energy heavy ion collision.\\
(3) The symmetry energy term affects the LMF's more than that of free nucleons. The $(N-Z)^2$ plays a crucial role \cite{21}. It has been studied that the isospin effects plays more important role in case of LMF's rather than free nucleons.\\
 From fig.2, it is clear that the minimum of fragment production is achieved at N=Z in case of FN's because they are produced in collision dynamics. But for LMF's the fragment production is nearly constant.\\
\begin{figure}
\includegraphics[scale=0.38]{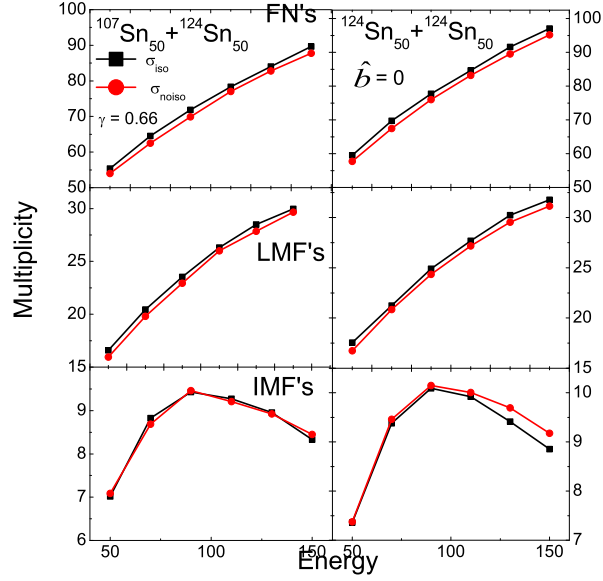}
\caption{\label{Fig:3} Multiplicity of free nucleons, LCP's and IMF's with energy at fixed scaled impact parameter for 
 $_{50}Sn^{124}+_{50}Sn^{124}$ and $_{50}Sn^{107}+_{50}Sn^{124}$.}
\end{figure}
Fig.\ref{Fig:3}, shows the variation of multiplicity of free nucleons, LCP's and IMF's with energy for central collision, for $_{50}Sn^{124}+_{50}Sn^{124}$ and $_{50}Sn^{107}+_{50}Sn^{124}$ for two different nucleon-nucleon cross-sections. It has been observed that multiplicity of free nucleons and LMF's increases with increase in energy. On the other hand, one can see a "rise and fall" in the multiplicity of IMF's; this behaviour is similar to the behaviour shown by Aladin group \cite{22}. Moreover, number of free nucleons and LMF's produced is very large as compared to IMF's this is because for central collisions, interactions are violent so large number of free nucleons and LMF's produced. It is clear from the figure that slope of the curve is steeper  in case of $_{50}Sn^{124}+_{50}Sn^{124}$  than $_{50}Sn^{107}+_{50}Sn^{124}$ and this theoretical observation is in agreement with the experimental observation of Sfienti et al.\cite{22}. This rise is due to the fact that in case of neutron rich system, heavy residues with low excitation energy will predominantly emit neutrons, a channel that is suppressed in case of neutron-poor nuclei. Here one can see the difference in the production of FN's, LMF's and IMF's due to different cross-sections. Since the proton number in both the cases is same but neutron number is different, so we expect some difference in the production. 
\begin{figure}
\includegraphics[scale=0.38]{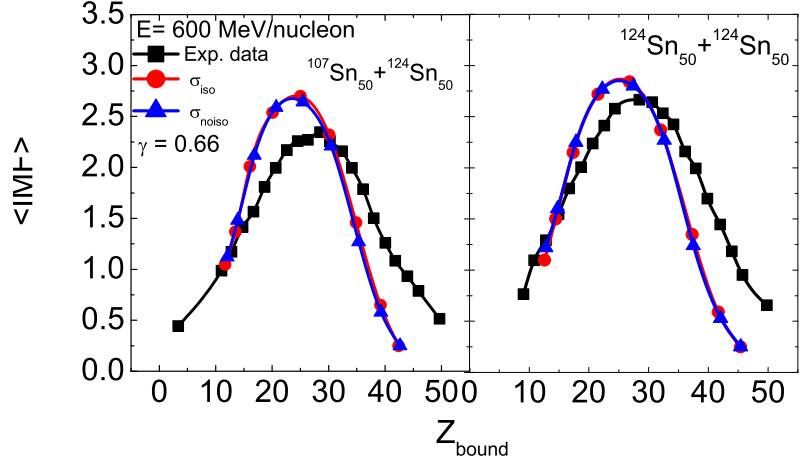}
\caption{\label{Fig:4} Multiplicity of IMF's as a function of $Z_{bound}$.}
\end{figure} 

In Fig.\ref{Fig:4}, 
we have shown IMF's as a function of $Z_{bound}$. The quantity $Z_{bound}$ is defined as sum of all atomic charges $Z_{i}$ 
of all fragments with $Z_{i}> 2$. Here we observe that at semi peripheral collisions the multiplicity of IMF's shows a peak because most of the spectator source does not take part in collision and large number of IMF's are observed. In case of central collision the collisions are violent so there few number of IMF's observed and for peripheral collisions very small portion of target and projectile overlap so again few number of IMF's observed most of the fragments goes out in heavy mass fragments (HMF's). In this way we get a clear "rise and fall" in multifragmentation emission. But the influence of $\sigma_{iso}$ and $\sigma_{noiso}$ is negligible here because IMF's are produed from the participant zone. It is observed that IMF's shows the agreement with data at low impact parameters but fails at intermediate impact parameters due to no acess to filters.\\ 
\section{\bf Summary}
By using isospin dependent quantum molecular dynamics model we have studied the role of isospin effects on fragmentation due to the collisions of charge asymmetric nuclei. Here calculations were carried out for $^{124}X_{n}+ ^{124}X_{n}$, where n varies from 47 to 59 and for $^{40}Y_{m}+ ^{40}Y_{m}$, where m varies from 14 to 23. It has been observed that isospin dependent cross-section shows its influence on fragmentation in the collision of neutron rich nuclei and there is a constant difference in the production of FN's and LMF's with $\sigma_{iso}$ and $\sigma_{noiso}$ for charge asymmetric nuclei.\\

\noindent

{\large{\bf Acknowledgment}}

This work has been supported by a grant from the university grant commission, Government of India [Grant No. 39-858/2010(SR)].\\

\end{document}